\documentclass[final,3p,times,twocolumn]{elsarticle}
\usepackage{amssymb}
\usepackage{amsmath}
\usepackage{lineno}
\usepackage{microtype} 
\usepackage[utf8x]{inputenx}
\journal{Nuclear Instruments and Methods in Physics Research A}
\begin{document}

\begin{frontmatter}


%%\ead[url]{home page}
 %%\address{Address\fnref{label3}}
%%\fntext[label3]{}

\title{Nano-machining, surface analysis and emittance measurements of a copper photocathode at SPARC\_LAB}

%%\address[label1]{}
\author[lnf]{J. Scifo}\ead{jessica.scifo@lnf.infn.it}
\author[lnf]{D. Alesini}
\author[lnf]{M.P. Anania}
\author[lnf]{M. Bellaveglia}
\author[lnf]{S. Bellucci}
\author[lnf]{A. Biagioni}
\author[lnf]{F. Bisesto}
\author[lnf]{F. Cardelli}
\author[lnf]{E. Chiadroni}
\author[roma2]{A. Cianchi}
\author[lnf]{G. Costa}
\author[lnf]{D. Di Giovenale}
\author[lnf]{G. Di Pirro}
\author[lnf]{R. Di Raddo}
\author[slac]{D. H. Dowell}
\author[lnf]{M. Ferrario}
\author[lnf]{A. Giribono}
\author[lecce]{A. Lorusso}
\author[lnf,chieti]{F. Micciulla}
\author[roma1]{A. Mostacci}
\author[roma1]{D. Passeri}
\author[lecce]{A. Perrone}
\author[lnf]{L. Piersanti}
\author[lnf]{R. Pompili}
\author[lnf]{V. Shpakov}
\author[lnf]{A. Stella}
\author[trieste]{M. Trov\`o}
\author[lnf]{F. Villa}

\address[lnf]{Laboratori Nazionali di Frascati, Via Enrico Fermi 40, 00044 Frascati (Roma), Italy}
\address[roma2]{Universit\`a di Roma "Tor Vergata" and INFN-Roma Tor Vergata, Via della Ricerca Scientifica 1, 00133 Rome, Italy}
\address[slac]{SLAC, Menlo Park, California 94025, USA}
\address[lecce]{Universit\`a del Salento, Dipartimento di Matematica e Fisica E. De Giorgi, INFN-Sezione di Lecce, 73100 Lecce, Italy}
\address[chieti]{Universit\`a degi Studi di Chieti e Pescara "G. D'Annunzio" , Dipartimento di Neuroscienze, Immaging e Scienze Cliniche Via dei Vestini, 33 66100 Chieti, Italy}
\address[roma1]{SBAI- Universit\`a di Roma ``La Sapienza'', via Antonio Scarpa, 24-00133 Roma, Italy}
\address[trieste]{ Elettra - Sincrotrone Trieste SCpA, S.S. 14 Km 163.5 in Area Science Park, 34149 Basovizza - Trieste, Italy}

\begin{abstract}
R\&D activity on Cu photocathodes is under development at the SPARC\_LAB test facility to fully characterize each stage of the photocathode ``life'' and to have a complete overview of the photoemission properties in high brightness photo-injectors. The nano(n)-machining process presented here consists in diamond milling, and blowing with dry nitrogen. This procedure reduces the roughness of the cathode surface and prevents surface contamination introduced by other techniques, such as polishing with diamond paste or the machining with oil. Both high roughness and surface contamination cause an increase of intrinsic emittance and consequently a reduction of the overall electron beam brightness.
To quantify these effects, we have characterized the photocathode surface in terms of roughness measurement, and morphology and chemical composition analysis by means of Scanning Electron Microscopy (SEM), Energy Dispersive Spectroscopy (EDS), and Atomic Force Microscopy (AFM) techniques.
The effects of n-machining on the electron beam quality have been also investigated through emittance measurements before and after the surface processing technique.
Finally, we present preliminary emittance studies of yttrium thin film on Cu photocathodes.

\end{abstract}


\end{frontmatter}

\section{Introduction}
\label{Introduction}
Most accelerator physics applications such as FEL (Free Electron Laser) and PWFA (Plasma Wake Field Acceleration), require high peak current and low emittance. A low emittance and high charge electron beam is needed for Inverse Compton Scattering sources, while, for the generation of THz signal high charge and high current electron beam are necessary.
All these requirements are specified in terms of high brightness of electron beam.
The improvement of brightness consists in an enhancement of the quantum efficiency, defined as the ratio of number of emitted electrons to the number of incident photons, and in a minimization of intrinsic emittance, that realizes the transverse momentum of the electrons emitted from the photocathode, in terms of their thermal energy~\cite{dowell2009quantum}. 
The source brightness is the highest brightness that beam can have. It can not be improved but only spoiled along the downstream accelerator. 
A photoinjector is the most popular among many and various types of schemes for high brightness electron beams generation and the performance of the photocathode is essential increasing the beam brightness~\cite{dowell2016sources}.
In the RF gun the photocathode is subject to surface modification and contamination due to laser, RF fields and low vacuum pressure.
A R\&D activity on photocathodes is under development at the SPARC\_LAB test facility~\cite{ferrario2013sparc_lab} to understand and characterize each stage of the photocathode's life cycle of production, installation and use and to improve photocathode performance of quantum efficiency and electron beam intrinsic emittance.




\section{Morphological and chemical analysis}
\label{Morphological and chemical analysis}
\begin{figure}[!h]
\begin{center}
\includegraphics[width=8cm]{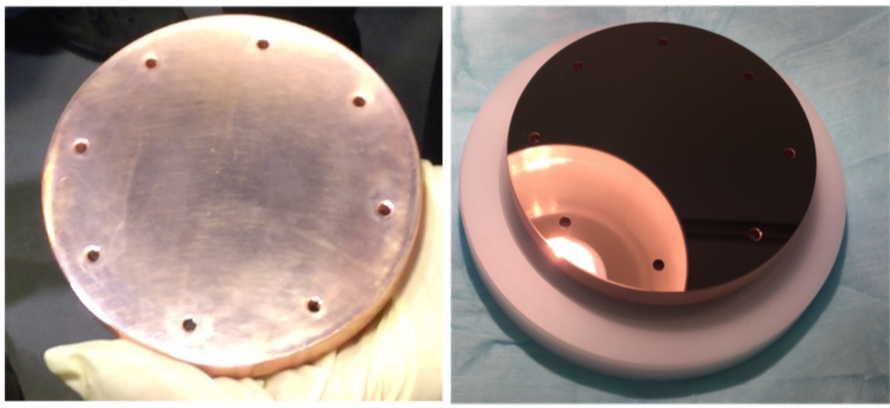}
\caption{Photocathode's surface. Left: surface before n-machining. Right: surface after n-machining. The machining removed about 100 $\mu$m of surface.}
\label{cathode1}
\end{center}
\end{figure}

The Cu photocathode surface has been machined by a german company, LT-ULTRA, by means of single cristal milling and clean with dry nitrogen. The machining has been done without the use of any oil or cooling fluid (dry machining).\\ Such a process has been defined as \textit{n-machining}. This procedure is useful to reduce roughness and to avoid surface contamination compared to other procedures for example the polishing with diamond paste or the machining with oil.
Before machining and after approximately 6 years of operation in the SPARC gun, the photocathodes surface appeared opaque due to surface oxidation, as shown on the left in Fig.~\ref{cathode1}.

\begin{figure}[h]
\includegraphics[width=8cm]{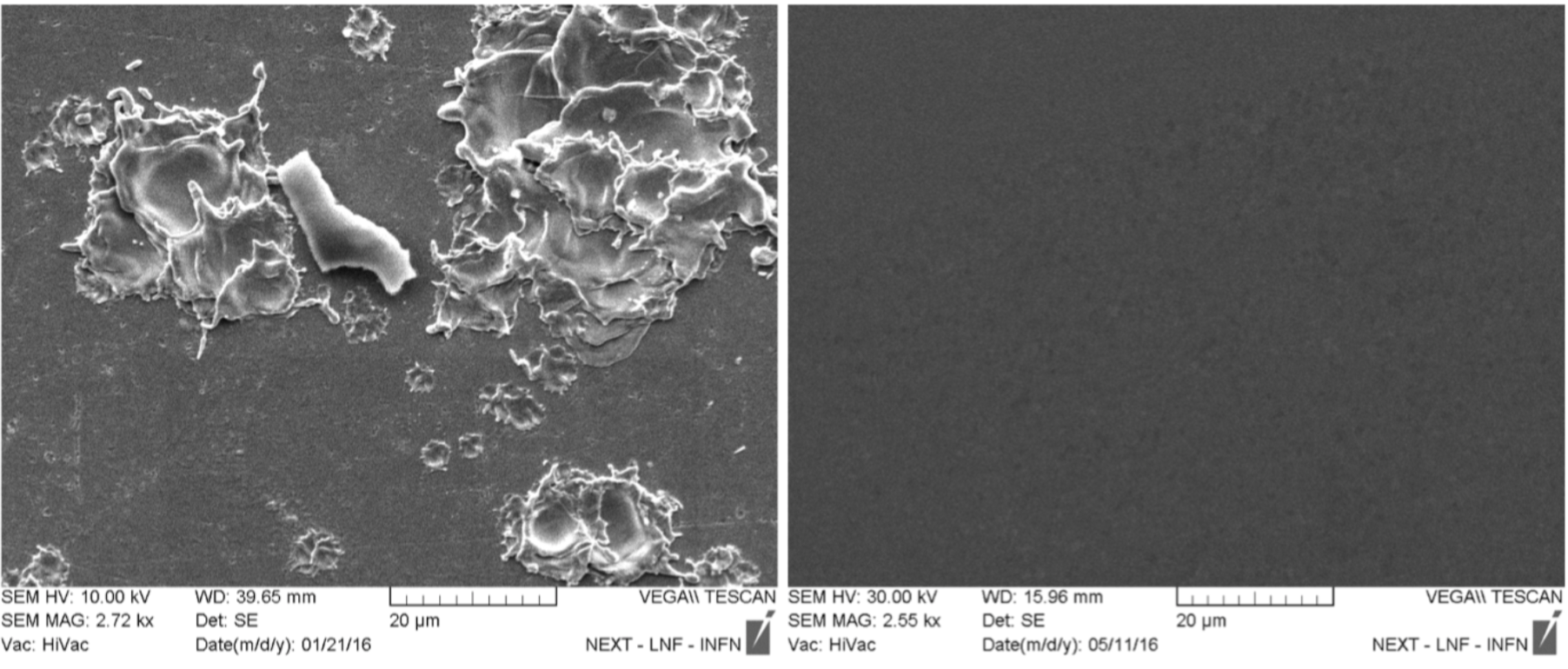}
\caption{Left: SEM analysis before n-machining. Right: SEM ana\-ly\-sis after n-machining at same resolution (20 $\mu$m/div).}
\label{old_new_cathode}
\end{figure}

\begin{figure}[!h]
\begin{center}
\includegraphics[width=8cm]{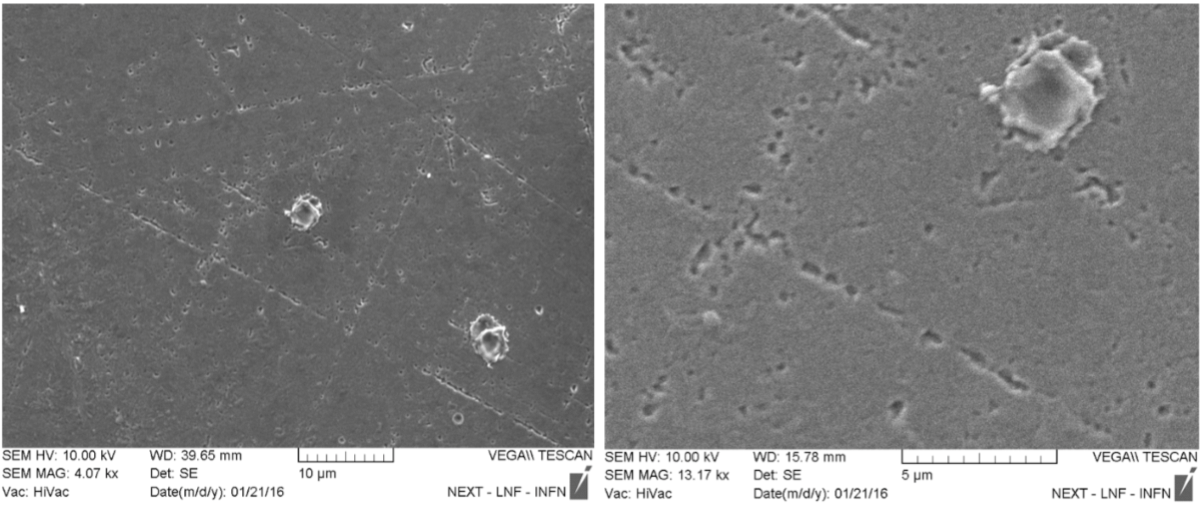}
\caption{SEM analysis before n-machining: on right a detail of the left picture is shown.}
\label{SEM_old}
\end{center}
\end{figure}

\begin{figure}[!h]
\centering
\includegraphics[width=8cm]{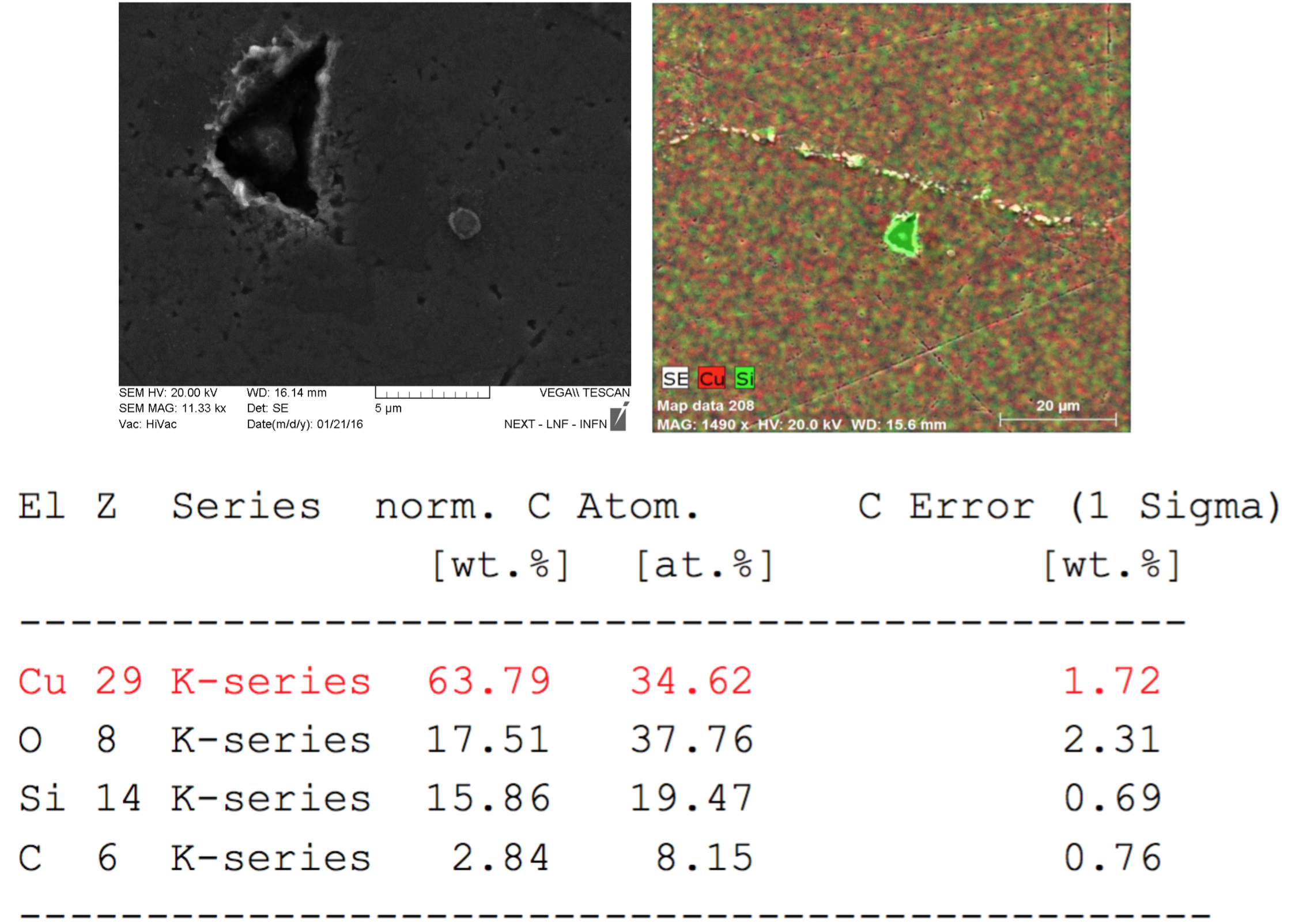}
\footnotesize
\caption{EDS analysis of the area before n-machining. Top:SEM image and mapping of presence of marked chemical elements of the area. Bottom: table of semiquantitative ana\-ly\-sis of surface composition. The third column shows the atomic shell of the element. The fourth, fifth and sixth columns show respectively the weight normalized percentage of the element, the atomic number normalized percentage of the element and its error. (Color on line).}
\label{EDSbefore}
\end{figure}

\begin{figure}[h]
\includegraphics[width=8cm]{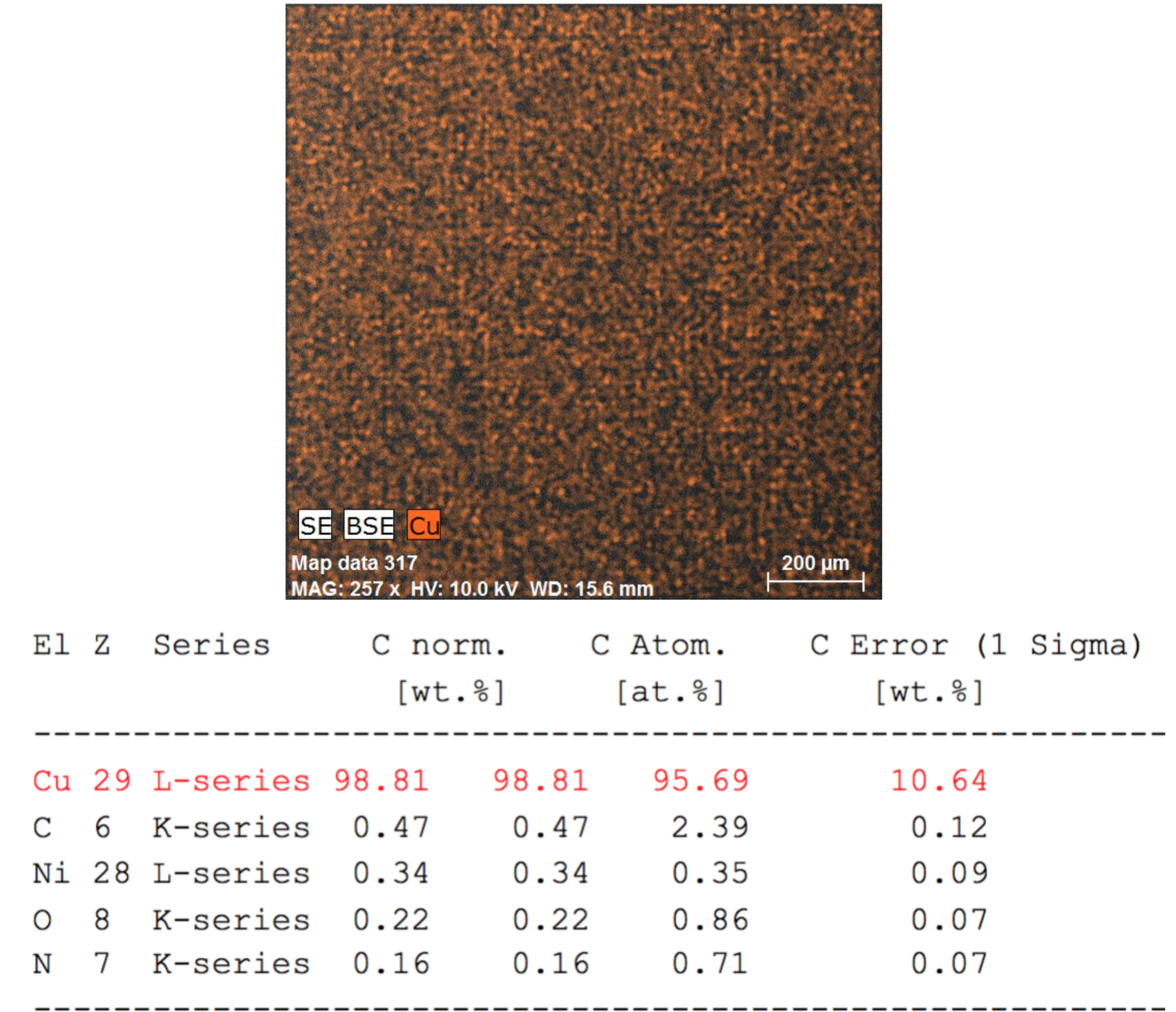}
\small
\caption{EDS analysis after n-machining. Left: mapping of presence of marked chemical elements of the area. Bottom: table of semiquantitative ana\-ly\-sis of surface composition. The third column shows the atomic shell of the element. The fourth, fifth and sixth columns show respectively the weight normalized percentage of the element, the atomic number normalized percentage of the element and its error.}
\label{EDSafter}
\end{figure}

Before and after machining the photocathode's surface has been analyzed by Scanning Electron Microscopy (SEM), Energy Dispersive Spectroscopy (EDS) to measure the chemical composition of surface~\cite{goldstein2017scanning} and by Atomic Force Microscopy (AFM)~\cite{yang2007application}.\\
The SEM analysis (on the left in Fig.~\ref{old_new_cathode}) shows a surface with many craters typical of RF field breakdowns and contaminants. On the right is the photocathode surface after n-machining. The n-machining has removed craters and any other impurities.
In Fig.\ref{SEM_old} we see other craters at two different resolutions. In addition to the craters, there are scratches: given the dimensions we can assume that they are due to the diamond paste used previously to polish the photocathode's surface.\\
Before n-machining, we have analyzed different areas of the surface with the EDS technique to determine the chemical composition. In the area shown in (Fig.\ref{EDSbefore}) we found traces of silicon in addition to copper, carbon and oxygen. Carbon and oxygen are due, respectively, to contamination and oxidation. We did not expect to find silicon but that was likely coming from diamond paste used for former polishing.\\
After machining, the analysis with the EDS shows a clean surface and purity about 99\%. The other elements are likely due to the contact with air before SEM and EDS analyses (Fig.~\ref{EDSafter}).\\

\section{Surface roughness induced emittance estimate}
\label{Surface roughness induced emittance estimate}
The AFM images has been analysed to estimate the photocathode surface roughness and its contribution to the emittance.
The photoca\-tho\-de surface roughness plays an important role on the beam emittance because it increases the transverse momentum of the emitted electron bunch. 
The surface roughness is represented by the rms roughness ($R_{q}$) that is defined as the root mean square of the profile height deviations from the mean line, recorded within the evaluation length L:
\begin{equation}
R_{q}=\sqrt{\frac{1}{L}\int_{0}^{L} Z(x)^2 dx}
\label{RQ} 
\end{equation}
where \textit {Z(x)} is the profile height function.
If we consider a sinusoidal surface, its shape can be modeled as $z=acos((2\pi/\lambda)x)$, where $a$ is the amplitude of the uneven surface and $\lambda$ is the wavelength of the surface fluctuation. The contribution to the emittance due to the surface roughness is then given by:~\cite{zhang2015analytical,xiang2007first}:
%\begin{figure}[h]
%\includegraphics[width=7.5cm]{2D_surf.png}
%\caption{2D sinusoidal surface.}
%\label{2Dsurf}
%\end{figure}
\begin{equation}
\varepsilon_{roughness}=\sigma_{x}\sqrt{\frac{e\pi^2a^2E_{rf}sin\Phi_{rf}}{2m_{0}c^2\lambda}}
\label{emitsemplice} 
\end{equation}
where $\sigma_{x}$ is the rms laser spot size, $e$ is the electron charge, $E_{rf}$ is the peak field applied, $\Phi_{rf}$ is the laser launch rf phase, $m_{0}$ is the electron invariant mass and $c$ is the speed of light.
We feel this evaluation is insufficient to give an accurate value for the surface roughness emittance because it does not account for the asymmetric details and range of spatial frequency of the photocathode's surface roughness. 
Therefore, the surface has been modelled using a Fourier series:
\begin{equation}
\varepsilon_{roughness}^2=\sum_{n=0}^{N-1}\varepsilon^2(a_{n},\lambda_{n})=\sigma_{x}^2{\frac{e\pi^2E_{rf}sin\Phi_{rf}}{2m_{0}c^2}}\sum_{n=0}^{N-1}\frac{a_{n}^2}{\lambda_{n}}.
\label{emitDowell} 
\end{equation}
This model assumes a 1D Fourier transform analysis of a line-out across the cathode, where the cathode surface is cosine-like and given by $z(x)=\sum_{n=0}^{N-1}a_{n}cos((2\pi/\lambda_{n})x)$. Applying Euler's formula, $e^{j\phi}=cos\phi+jsin\phi$, the 1D Fourier transform is:
\begin{equation}
Re[F(l)]=\frac{1}{N}\sum_{n=0}^{N-1}Re[f(n)]cos2\pi\frac{n}{N}l
\label{1DDFT} 
\end{equation}
Assuming the line-out length is the photocathode length $L_{cathode}$, and comparing Eqn.~\ref{1DDFT} with the equation for $z(x)$, it is possible to write the following connections between the 1D Fourier transform and the emittance theory coefficients and wavelengths as: $z(x)\leftrightarrow Re[F(l)]$, $a_{n}=Re[f(n)]$, $k_{n}=2\pi\frac{n}{L_{cathode}}$, $\lambda_{n}=\frac{L_{cathoden}}{n}$.
Since the Fourier series is orthonormal, there are no cross terms between different spatial frequencies. This allows us to write the surface roughness induced emittance as the quadratic sum of the emittance at each frequency(~\ref{emitDowell}). The Fourier transforms have been calculated using ten 1D-line profiles~\cite{feng2017near} from the AFM images before and after n-machining. The final surface roughness induced emittance is evaluated at the experimental paramaters: $E_{rf}=97 MV/m$, $\Phi_{rf}=30^\circ$ and $\sigma_{x}=0.3mm$. 
Before n-machining the estimate is $\varepsilon_{roughness}\approx0.04$ mmmrad whereas after n-machining it is $\varepsilon_{roughness}\approx0.004$ mmmrad.
\begin{figure}[h]
\includegraphics[width=8cm]{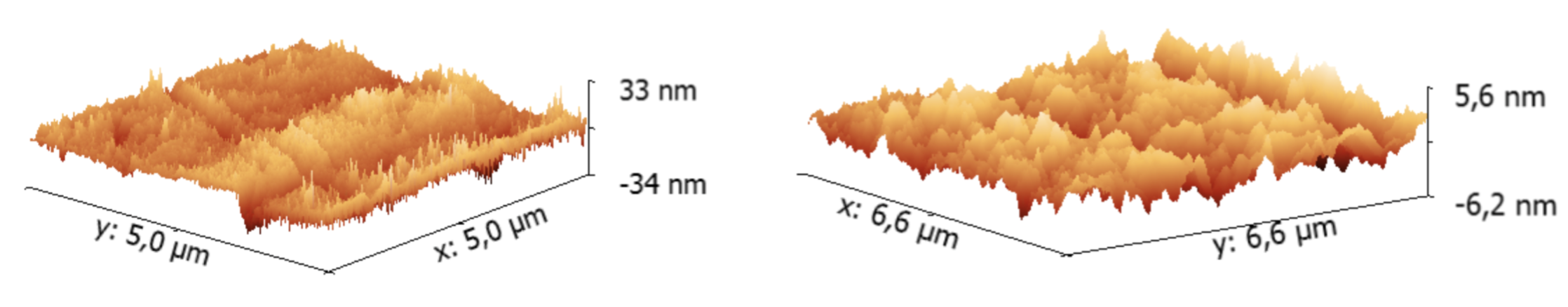}
\caption{AFM 3D images before (left) and after (right) the n-machining.}
\label{AFM}
\end{figure}
From the AFM analysis (Fig.~\ref{AFM}) we have found a rms roughness $R_{q}=5.7 nm$ for the photocathode's surface before n- machining and $R_{q}=1.5 nm$ after n-machining.
%In table~\ref{surface roughness induced emittance} the surface roughness induced emittance estimates are reported.
%\%begin{table}[!h]
%\centering
%\caption{\small{The surface roughness induced emittance estimates calculated at experimental paramaters: $E_{rf}=97 MV/m$, $\Phi_{rf}=30^\circ$ and $%\sigma_{x}=0.3mm$.}}
%\label{surface roughness induced emittance}
%\vspace*{1mm} 
%\begin{tabular}{ c | c }
%\hline \hline
%Before n-machining & After n-machining\\ \hline
%$\varepsilon\textsubscript{total}\approx0.04$ mmmrad & $\varepsilon\textsubscript{total}\approx0.004$ mmmrad \\ 
%\hline \hline
%\end{tabular}
%\end{table}
\section{Beam emittance measurements}
\label{Beam emittance measurements}
\subsection{Cu photocathode}
The experimental beam emittance has been measured using the solenoid scan technique. Measurements of beam size on a YAG screen, placed $1.181 m$ downstream from the solenoid, have been acquired for different solenoid fields.
In a solenoid scan beam size measurements for at least three different solenoid settings are required in order to solve for the three independent unknown parameters ($\langle x\textsubscript{0} ^2\rangle$, $\langle x_{0} x'_{0}\rangle$ and $\langle {x'^2_{0}}\rangle$). Such a system is overdetermined and it can be solved by the standard technique of the $\chi^2$ minimization~\cite{mostacci2012chromatic,graves2001duvfel}: 
\begin{equation}
\langle x\textsubscript{(i)} \rangle^2=R^{(i)^2}_{11}\langle x\textsubscript{0} ^2\rangle+2R^{(i)}_{11}R^{(i)}_{12}\langle x_{0} x'_{0}\rangle+R^{(i)^2}_{12}\langle {x'^2_{0}}\rangle
\end{equation}
where (${i}$) is the measurements number and the coefficients $R_{11}$ and $R_{12}$ are the elements of the beam line transfer matrix.\\
The normalized emittance (RF emittance, space charge emittance, intrinsic emittance and surface roughness induced emittance combined in quadrature) has been computed at the entrance of the gun solenoid:
\begin{equation}
\varepsilon_{nx,rms}=\gamma\beta\sqrt{\langle x\textsubscript{0} ^2\rangle\langle {x'^2_{0}}\rangle-\langle x_{0} x'_{0}\rangle^2}.
\end{equation}
\begin{figure}[!h]
\centering
\includegraphics[width=7cm]{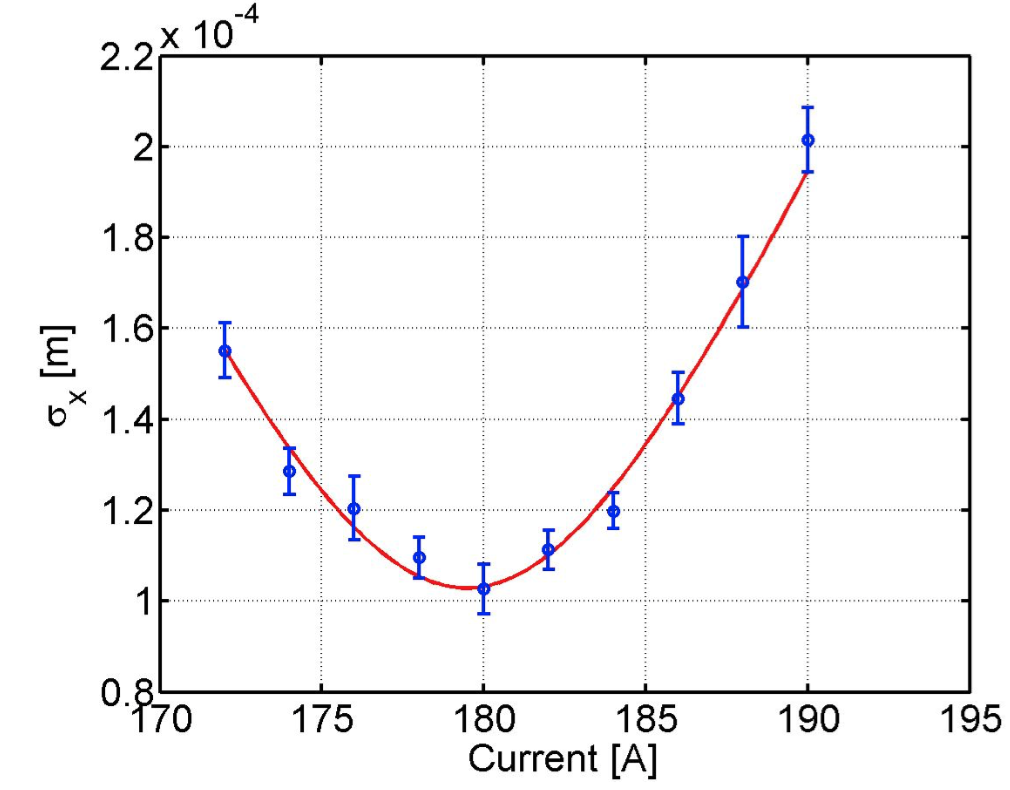}
\caption{Typical example of a solenoid scan performed. The plot shows the beam size versus the soleinod current. Experimental data are reported with blue dots, fit is reported with a solid red line. The bunch charge is $\sim7$ pC. The other relevant parameters are reported in tab.~\ref{Tab_beam parameters_Cu scaling with bunch charge}.}
\label{SOLSCAN}
\end{figure}
Before and after n-machining we have performed measurements by varying beam parameters. At the same fixed beam parameters, bunch charge $\sim6$ pC, laser pulse length  = 5 ps (FWHM), laser spot size (rms)$\sim0.3$ mm, E\textsubscript{rf} = $84$ MV/m and working rf phase =$30^\circ$ we have retrieved the beam emittance before and after n-machining and compared the results, as shown in tab.~\ref{Tab_beam parameters_Cu before and after n-machining}.\\
\begin{table}[!h] \footnotesize
 %\small 
\caption{\small{Beam emittance values before and after n-machining.  The used parameters are: bunch charge $\sim6$pC, laser pulse length  = 5 ps (FWHM), E\textsubscript{rf} = $84$ MV/m.}}
\label{Tab_beam parameters_Cu before and after n-machining}
\vspace*{1mm} 
\begin{tabular}{ c | c }
\hline \hline
Before n-machining & After n-machining\\ \hline
$\varepsilon\textsubscript{nx,rms}$= $0.24\pm0.04$ mm-mrad & $\varepsilon\textsubscript{nx,rms}$= $0.13\pm0.02$ mm-mrad \\ 
\hline
 $\varepsilon\textsubscript{ny,rms}$=$0.28\pm0.04$ mm-mrad & $\varepsilon\textsubscript{ny,rms}$= $0.15\pm0.02$ mm-mrad\\
\hline \hline
\end{tabular}
\end{table}
After n-machining we have performed measurements by varying bunch charge. The parameters are reported in tab.~\ref{Tab_beam parameters_Cu scaling with bunch charge}.
\begin{table}[h]
\centering
\small
\caption{\small{Beam parameters of the emittance measurements corresponding to Fig.~\ref{emit_charge_sparc}}}
\label{Tab_beam parameters_Cu scaling with bunch charge}
\vspace*{1mm} 
\begin{tabular}{ c  c }
\hline \hline
Parameters & Value\\ \hline
E\textsubscript{rf}   & $97$ MV/m\\ 
Working rf phase  & $30^\circ$\\
Laser pulse length     & $5$ ps - FWHM (Gaussian profile)\\ 
Energy at the gun exit   & $4.53 \pm 0.05$ MeV\\ 
Laser spot size (rms) & $\sim0.3$ mm (Flat top profile) \\
\hline \hline
\end{tabular}
\end{table}
\begin{figure}[!h]
\begin{center}
\centering
\includegraphics[width=9cm]{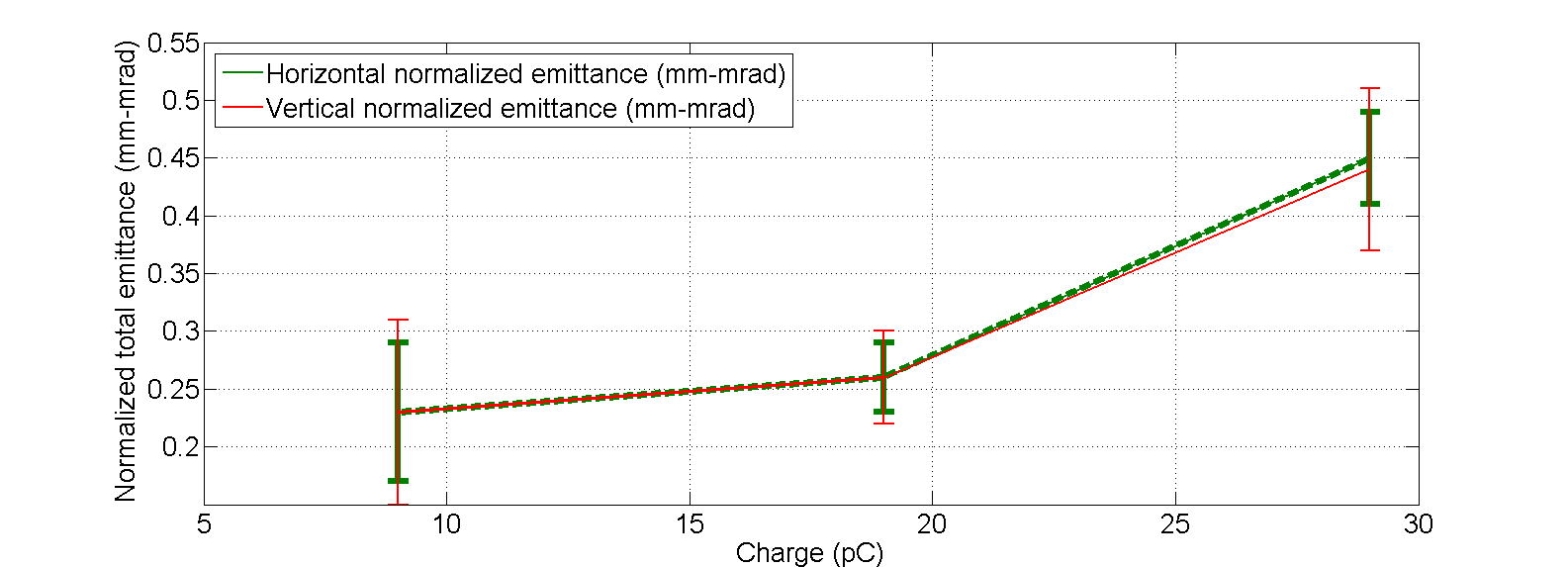}
\caption{ Plot of total horizontal (green line) and vertical (red line) normalized beam emittance vs bunch charge for Cu photocathode(Color on line).}
\label{emit_charge_sparc}
\end{center}
\end{figure}
%\vspace*{1mm} 
The results in Fig.~\ref{emit_charge_sparc} show that after $\sim19$ pC the emittance growth due to the space charge becomes significant.
\subsection{Preliminary emittance studies of yttrium thin film on Cu photocathodes}
In this section we outline the findings of the beam emittance preliminary study for a yttrium thin film (the thickness of yttrium film is 1.2$\mu$m) on Cu photocathode~\cite{cultrera2010photoemission, lorusso2011detailed,lorusso2017pulsed}.
We have performed measurements by varying bunch charge, at Cavity Test Facility (CTF) at Elettra - Sincrotrone Trieste~\cite{Allaria:ig5031}, using the solenoid scan technique.
The used parameters are reported in tab.~\ref{Tab_beam parameters_Y scaling with bunch charge}.\\
\begin{table}[!h]
\centering
\small
\caption{\small{Beam parameters of emittance measurements corresponding to Fig.~\ref{Fermi}}}
\label{Tab_beam parameters_Y scaling with bunch charge}
\vspace*{1mm} 
\begin{tabular}{ c  c }
\hline \hline
Parameters & Value\\ \hline
E\textsubscript{rf}   & $110$ MV/m\\ 
Working rf phase  & $30^\circ$\\
Laser pulse length     & $5.3$ ps - FWHM (Gaussian profile)\\ 
Energy at the gun exit   & $4.66$ MeV\\ 
Laser spot size (rms) & $\sim0.11$ mm (Gaussian profile) \\
\hline \hline
\end{tabular}
\end{table}
\begin{figure}[!h]
\begin{center}
\includegraphics[width=8cm]{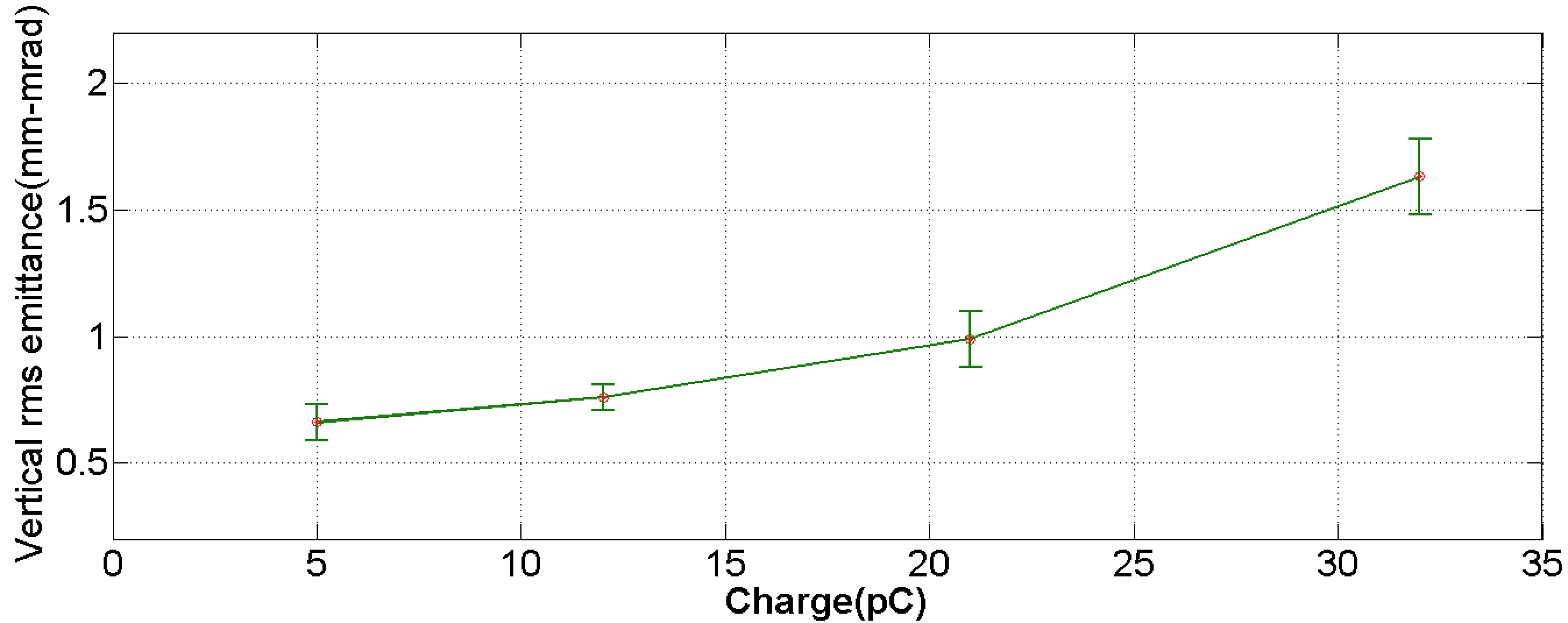}
\caption{Plot of normalized beam emittance vs bunch charge for Y thin film on Cu substrate photocathode(Color on line).}
\label{Fermi}
\end{center}
\end{figure}
The results in Fig.~\ref{Fermi} show that above $\sim12$ pC the emittance growth ($0.76\pm0.05$) mmmrad due to the space charge becomes significant.
We note that for the yttrium photocathode the beam emittance is approximately 3-times larger than Cu photocathode. 
The high emittance value is probably due to the high value of ($h\nu-\Phi_y$) in the intrinsic emittance formula~\cite{dowell2009quantum, graves2001measurement}. The yttrium work function is indeed $\Phi_y=3$ eV but the measurements have been performed with $\lambda_{laser}=262$ nm ($h\nu=4.73$ eV). For the yttrium $h\nu-\Phi_y=(4.73-3)eV=1.73eV$ (with this value the theoretical normalized emittance is about $\varepsilon_N$=$1$ mm-mrad/mm ) which is about 10 times larger than for the copper. The copper work function is $\Phi_y\sim4.59$eV and consequently $h\nu-\Phi_y=(4.73-4.59)eV\sim0.2eV$.
\section{Conclusions}
To produce high brightness electron beam the dry machining is a better procedure with respect the use of diamond paste or oil. With this kind of machining 
a roughness of about 2nm has been achieved, typical of monocrystalline Cu photocathode. In addition the analysis of the Fourier transforms showed a surface roughness induced emittance estimate after n-machining a factor 10 smaller than the surface roughness induced emittance estimate before n-machining . 
The preliminary results about yttrium photocathode show us that, among others hypotheses, the difference in emittance compared with copper is due to the opportunity to have linear electron photoemission with visible radiation ($\lambda_{laser}\sim400$ nm) instead UV radiation ($\lambda_{laser}=266$ nm). Further studies are foreseen with visible light.
\section*{Acknowledgments}
This work was supported by the European Union's Horizon 2020 research and innovation program under grant agreement No. 653782.
\section*{References}
\bibliographystyle{elsarticle-num}
\bibliography{biblio}
\end{document}